\documentclass[12pt]{iopart}
\usepackage{epsfig}
\usepackage{graphicx}

\def\NN{\mbox{\tiny\rm NN}}
\def\be{\begin{equation}}
\def\ee{\end{equation}}
\def\bea{\begin{eqnarray}}
\def\eea{\end{eqnarray}}

\begin{document}
\title[Indication of quark deconfinement ...]
{{\bf  Indication of quark deconfinement and evidence for a Hubble flow in 130 
and 200 GeV Au+Au collisions} }
\author{M. Csan\'ad$^{\, 1}$, T. Cs\"org\H{o}$^{\, 2}$, B.
L\"orstad$^{\, 3}$ and A. Ster$^{\, 2}$}
\address{$^1$Dept. Atomic Phys., ELTE, H-1117 Budapest,
 	P\'azm\'any P. 1/a, Hungary}
\address{$^2$MTA KFKI RMKI, H - 1525 Budapest 114, P.O.Box 49, Hungary}
\address{$^3$Dept. Physics, University of Lund, S - 22362 Lund, Sweden}
\begin{abstract}
Buda-Lund hydro model fits are compared 
to BRAHMS, PHENIX, PHOBOS and STAR data on
identified particle spectra, two-particle Bose-Einstein or
HBT correlations, charged particle pseudorapidity
distributions and pseudorapidity as well as $p_t$ dependent elliptic
flow in $\sqrt{s_{\NN}} = 130$ and $200$ GeV Au+Au collisions at RHIC.
Preliminary results indicate that 7/8-th of the particle
emitting volume is  rather cold, 
with   surface temperature of 105 MeV, 
but the temperature has a distribution and  
the most central 1/8-th of the volume is 
superheated to $T(x) > T_c = 172 \pm 3$ MeV.
\end{abstract}

%Uncomment for PACS numbers title message
%\PACS{24.10.Nz, 25.75.-q, 25.75.Nq, 25.75.Gz}

% Uncomment for Submitted to journal title message
%\submitto{\JPA}

% Comment out if separate title page not required
%\maketitle

\section{Introduction}
The Buda-Lund hydro model~\cite{Csorgo:1995bi} is successful
 in describing the BRAHMS, PHENIX, PHOBOS
and STAR data on identified single particle spectra and the transverse mass
dependent Bose-Einstein or HBT radii as well as the pseudorapidity distribution
of charged particles in Au + Au collisions
both at $\sqrt{s_{\NN}} = 130 $ GeV~\cite{ster-ismd03} and
at $\sqrt{s_{\NN}} = 200 $ GeV~\cite{mate-warsaw03}.
Recently, Fodor and Katz calculated the phase diagram
of lattice QCD at finite net baryon density~\cite{Fodor:2001pe}.
Their results, obtained with light quark masses
four times heavier than the physical value,
indicated that in the $0 \le \mu_B \le 300$ MeV region the transition
from confined to deconfined matter is not a first or second order phase-transition, but a cross-over
with a nearly constant critical temperature,  $T_c = 172 \pm 3$ MeV.
The result of the Buda-Lund fits to Au+Au data,
both at $\sqrt{s_{\NN}} = 130 $ 
and $200 $ GeV,
indicate the existence of a very hot region. The temperature distribution
$T(x)$ of this region is characterized  with a central temperature $T_0$,
found to be greater than the critical value calculated from lattice QCD: 
$T_0 > T_c$~\cite{mate-ell1}.
The Buda-Lund fits thus indicate quark deconfinement
in  Au + Au collisions at RHIC.
The observation of a superheated center in Au+Au collisions at RHIC is
confirmed by the analysis of
$p_t$ and $\eta$ dependence of the elliptic flow~\cite{mate-ell1}, 
measured by the PHENIX~\cite{PHENIX-v2-id} 
and PHOBOS collaborations~\cite{PHOBOS-v2,PHOBOS-v2-qm02}.
A similar analysis of Pb+Pb collisions at CERN SPS energies 
yields central temperatures lower than the 
critical value, $T_0 < T_c$~\cite{ster-qm99,cs-rev}.

Here we summarize the Buda-Lund fit results as detailed in
refs.~\cite{ster-ismd03,mate-warsaw03,mate-ell1,ster-qm99}.
See these papers for definitions and discussion of the
 results as well as more detailed references.

\section{Buda-Lund fit results to central Au+Au data at
$\sqrt{s_{\NN}}= 130$ and $200$ GeV }
\begin{figure}[ht]
\begin{center}
\vspace{-0.5cm}
\begin{center}
\includegraphics[width=2.2in]{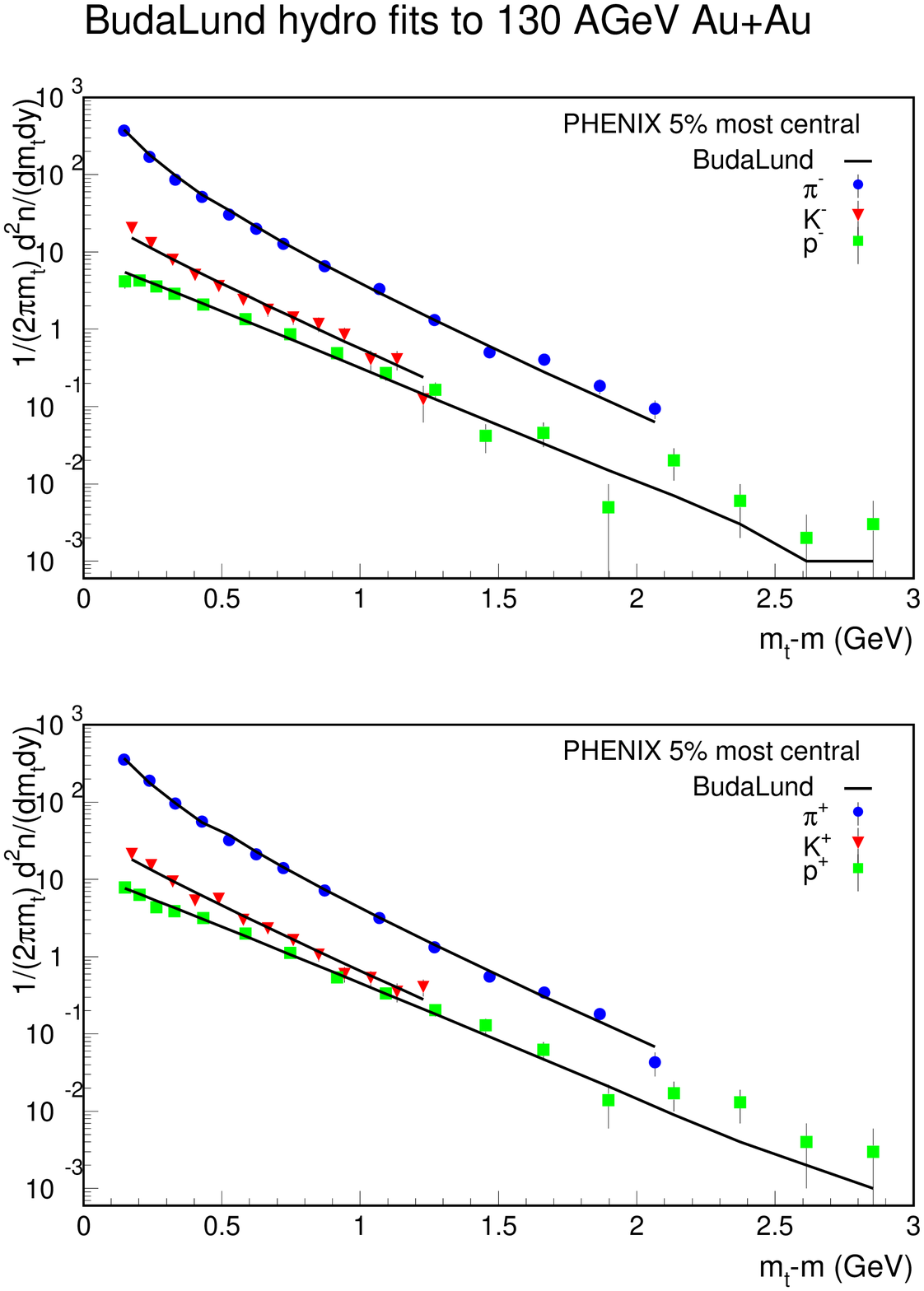}
\includegraphics[width=2.2in]{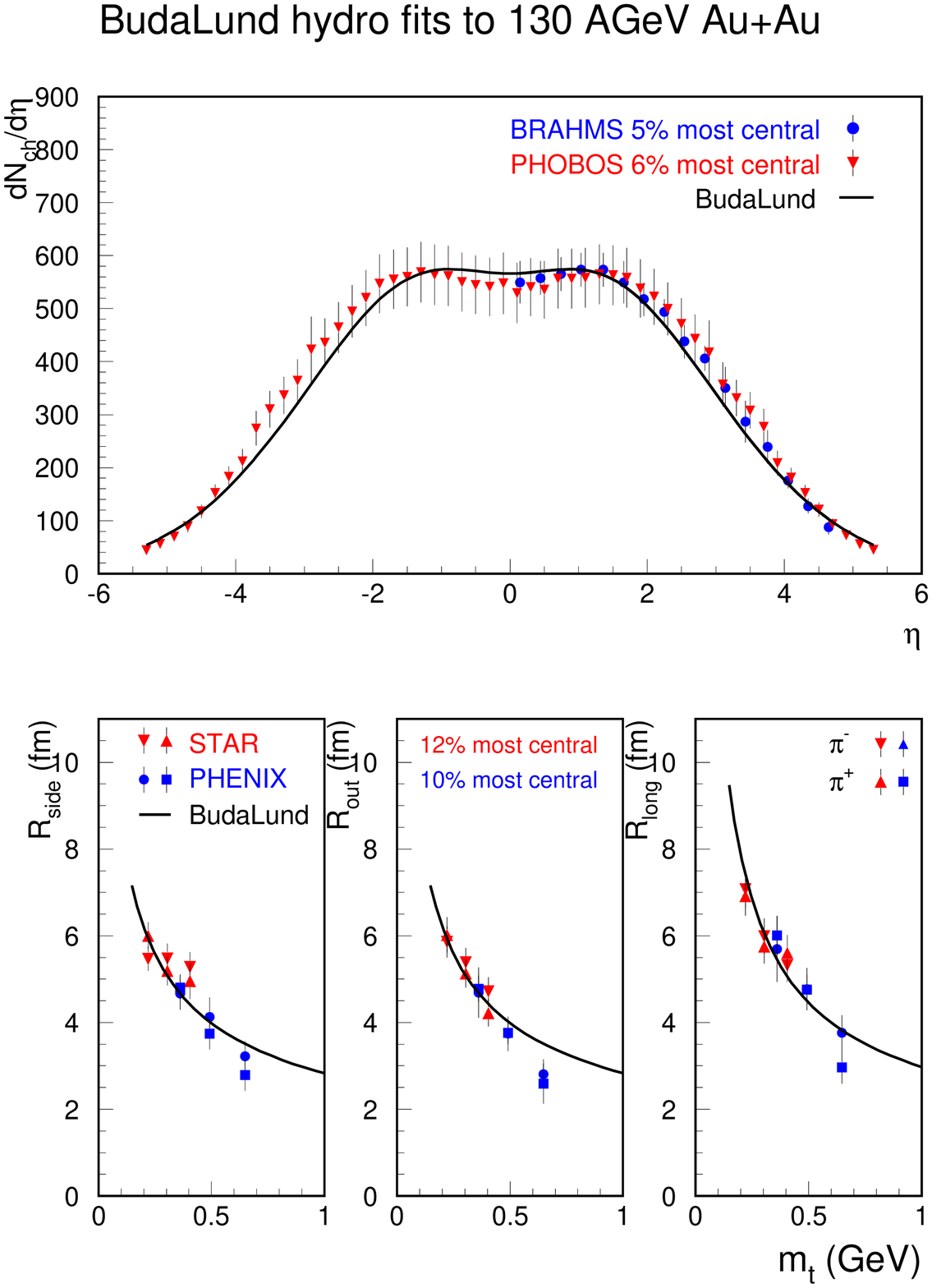} \\
	\includegraphics[width=2.2in]{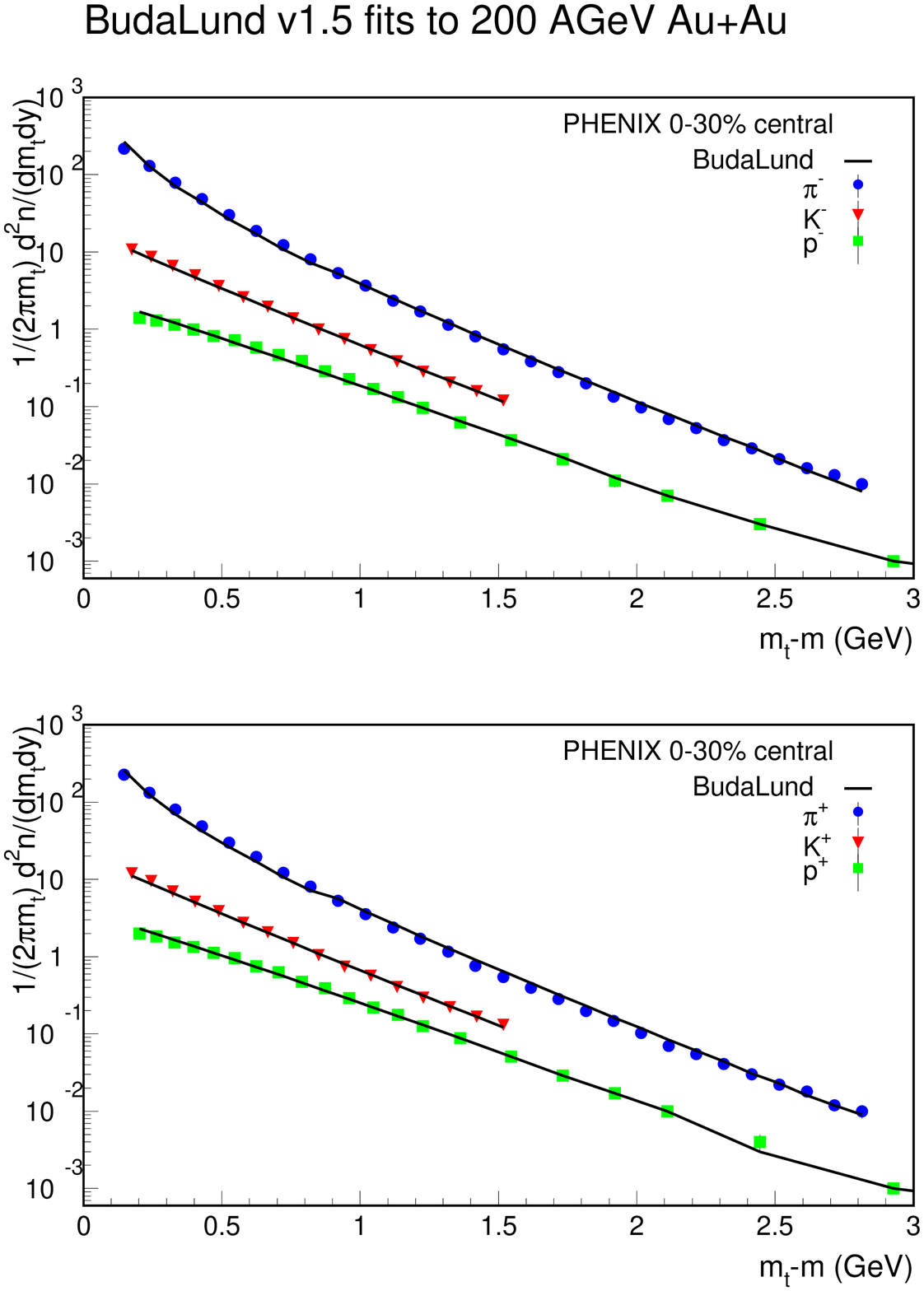}
	\includegraphics[width=2.2in]{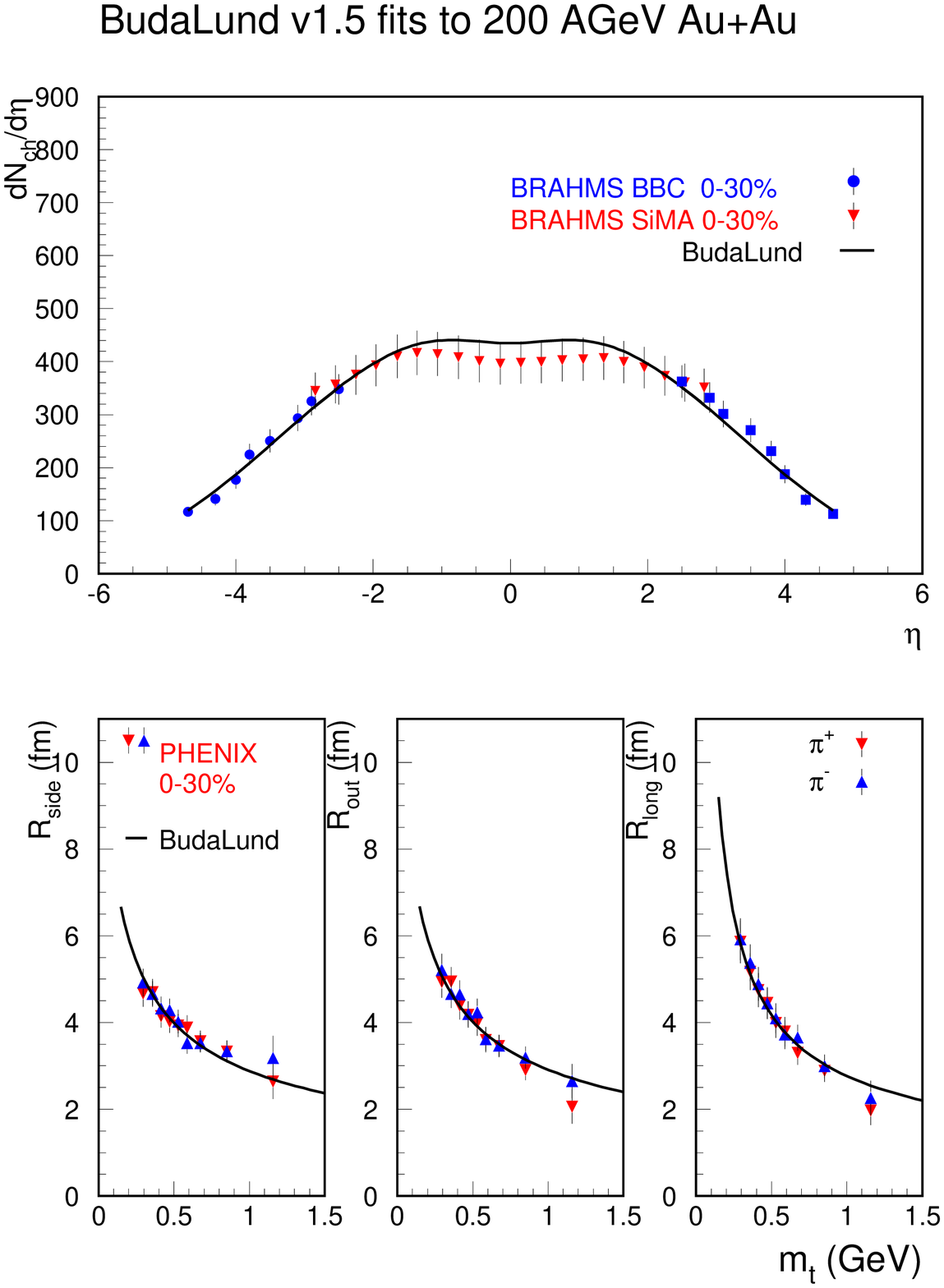}
\end{center}
\caption[*]{
\label{fig:130-spectra}{\small The upper four panels 
show a  simultaneous Buda-Lund 
fit  to 0-5(6)~\% central Au+Au data at $\sqrt{s_{NN}} = 130$ GeV,
refs.~\cite{Bearden:2001xw,Adcox:2001mf,Adcox:2002uc,Back:2001bq,Adler:2001zd}.
 The lower four  panels show similar fits to 0-30 \% central Au+Au data at
$\sqrt{s_{\NN}} = 200$ GeV, refs.~\cite{Bearden:2001qq,Adler:2003cb,Adler:2004rq}.
The fit parameters are summarized in Table 1.}
}
\end{center}
\vspace{-2truemm}
\end{figure}
Analyzing the fit parameters (Table 1) we find that $T_0 > T_c$
by more than $5 \sigma$ in case of the $0-5(6) \%$ most central Au+Au data at
$\sqrt{s_{NN}} = 130 \textrm{ GeV}$. We interpret this as an indication
of quark deconfinement. 
In case of the less central ($0-30 \%$) Au+Au data 
at $\sqrt{s_{NN}} = 200 \textrm{ GeV}$, 
with an improved analysis we find $T_0 > T_c$ by $2 \sigma$,
not a significant difference. 
We interpret this as a possible hint for quark deconfinement. 
In both cases, the flow profile within errors coincides with the three-dimensional Hubble flow, $u^\mu = x^\mu/\tau$.
For details, see refs.~\cite{cs-rev,Csorgo-blhome}. 
For similar results, % of the Cracow and the blastwave hydro models, 
see refs.~\cite{blastwave,cracow-sp,cracow-hbt}.

\section{Buda-Lund results for the elliptic flow in Au+Au data at RHIC}
\begin{figure}[ht]
\begin{center}
%\includegraphics[width = 2.7in]{v2pt.eps}
%\includegraphics[width = 2.7in]{v2eta.eps} \\
%\includegraphics[height = 2.7in, angle = 270]{vnpt.eps}
%\includegraphics[height = 2.7in, angle = 270]{vneta.eps}
%\end{center}
\includegraphics[width = 2.7in]{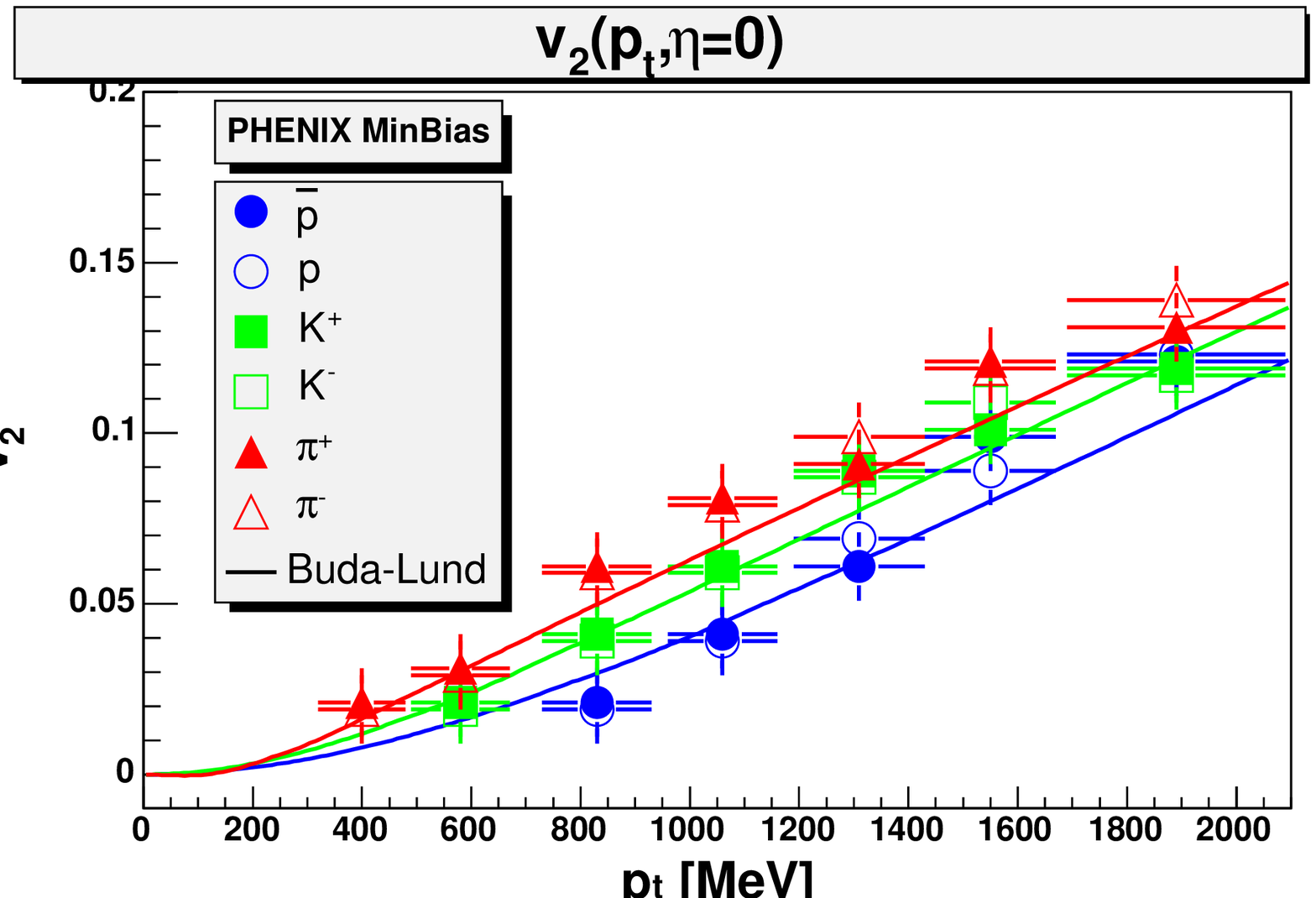}
\includegraphics[width = 2.7in]{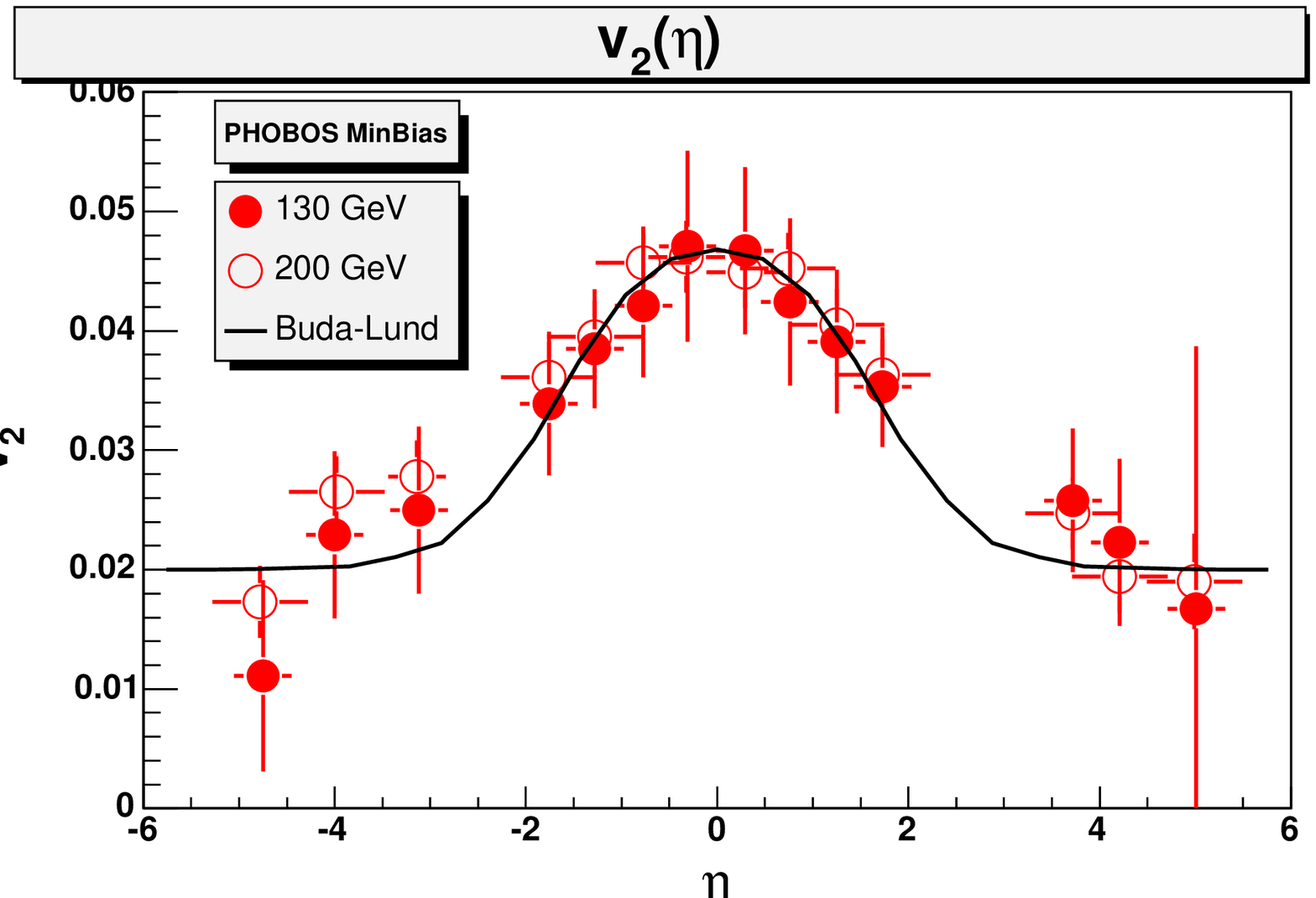} \\
\includegraphics[width = 2.7in]{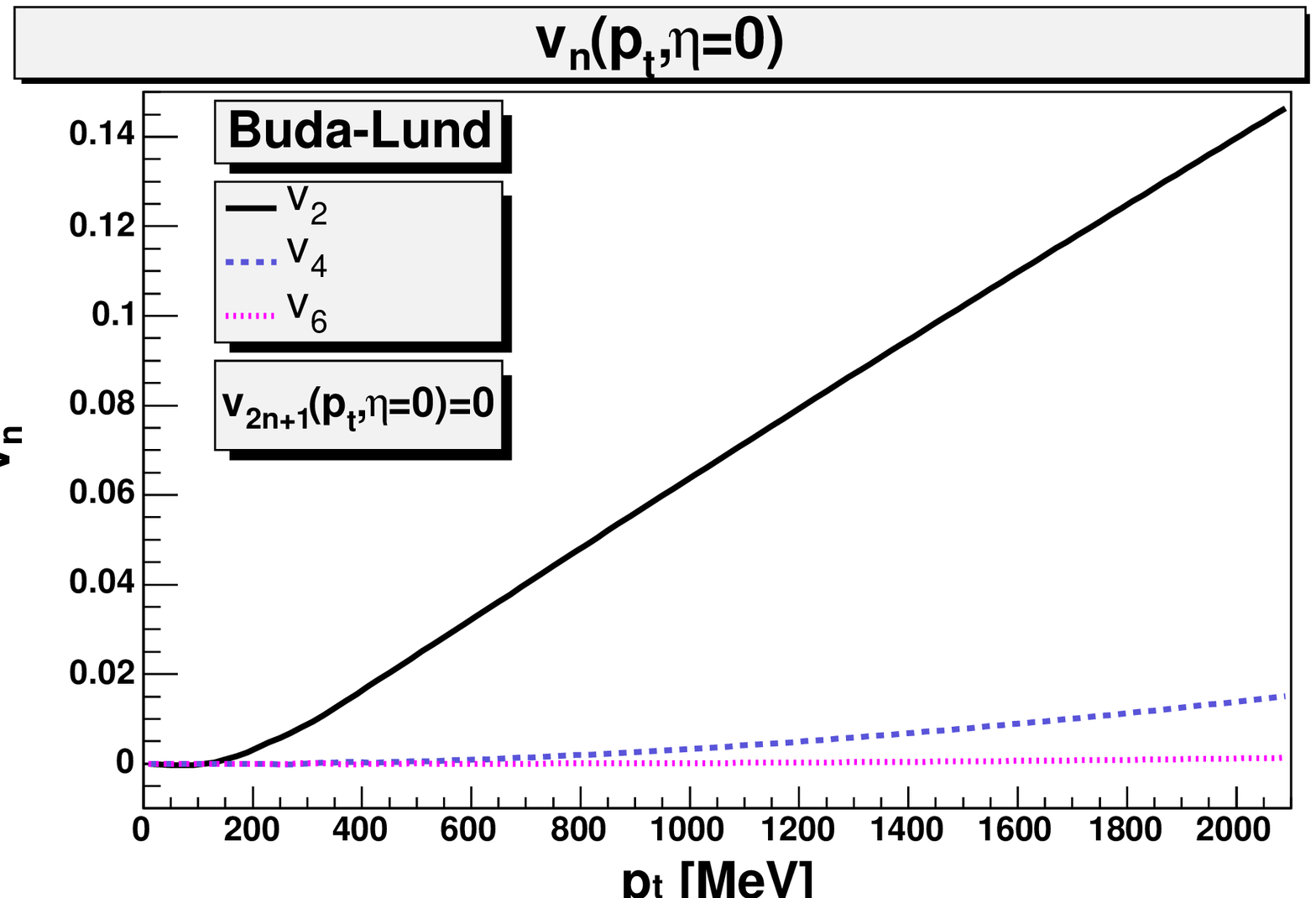}
\includegraphics[width = 2.7in]{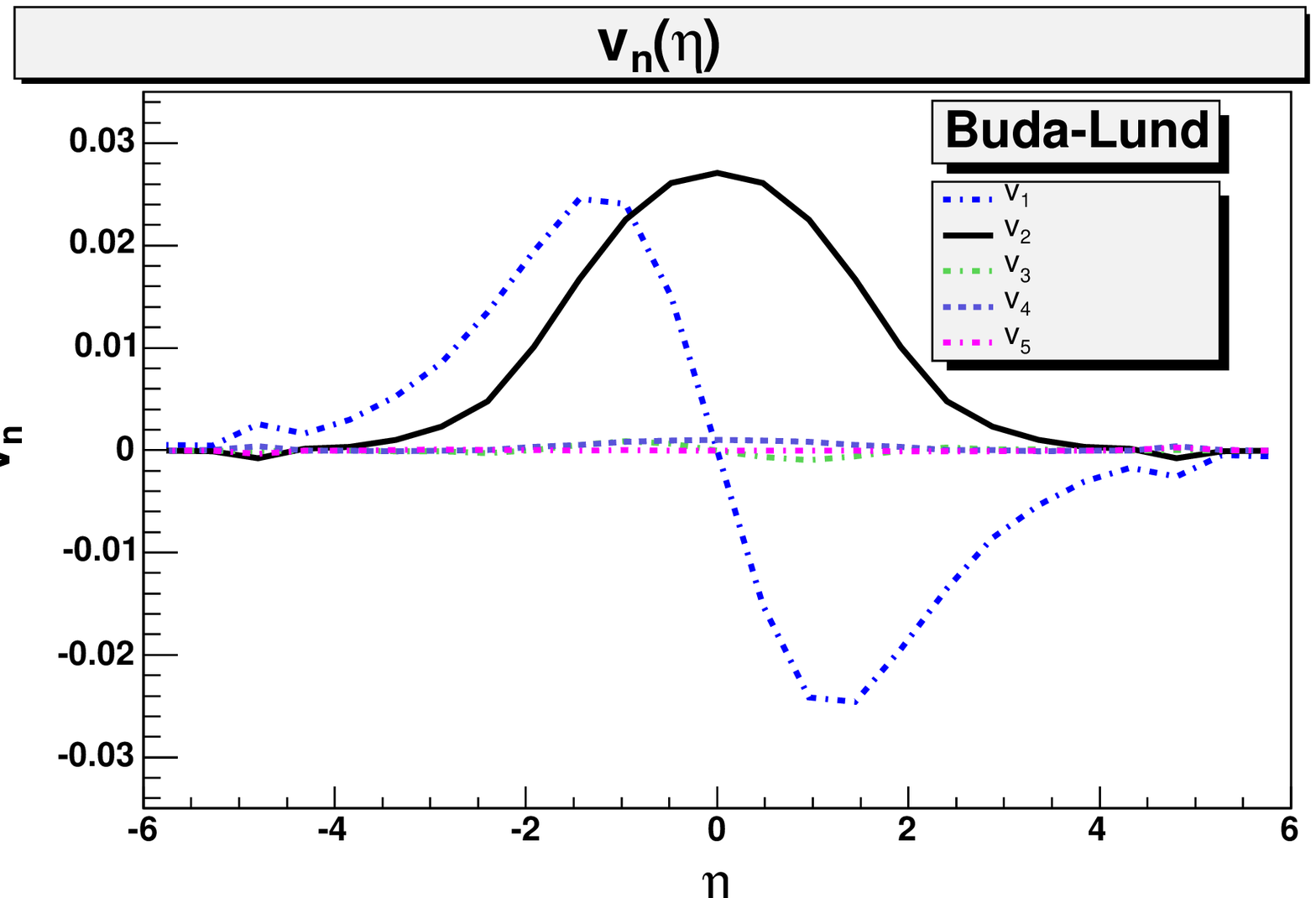}
\end{center}
\caption{In the top panels, the ellipsoidally symmetric Buda-Lund hydro
calculations are compared to PHENIX identified $v_2(p_t)$
data at 200 GeV~\cite{PHENIX-v2-id} and PHOBOS  $v_2(\eta)$ data at 130 and 200
GeV~\cite{PHOBOS-v2,PHOBOS-v2-qm02}\label{fig:v2}. On the top right panel, we added a constant non-flow parameter of 0.02 to the
calculated values of $v_2$. In the lower panels, the flow coefficients, $v_n$-s are shown for $n=1$ ...
$6$ as a function of transverse momentum at midrapidity, as well as a function
of pseudorapidity, with integrated $p_t$.}
\end{figure}

The model is evaluated in all panels with the following parameters:
$T_0=210\textrm{ MeV}$, $T_s=105\textrm{ MeV}$, $X_f=8.6$
fm, $Y_f=10.5$ fm, $Z_f=17.5$ fm, $\dot X_f=0.57$, $\dot Y_f=0.45$, $\dot
Z_f=2.4$, $\tau_0 = 7$ fm/c, $\vartheta=0.09$, $\mu_{0,\pi}=70\textrm{ MeV}$,
$\mu_{0,K}=210 \textrm{ MeV}$ and $\mu_{0,p}=315 \textrm{ MeV}$.
$T_0$ is the temperature of the center, while $T_s$ is that of the surface,
$(X_f,Y_f,Z_f)$ are the principal axes and $(\dot X_f,\dot Y_f, \dot Z_f)$
the principal expansion rates of the ellipsoid
at the freeze-out proper-time $\tau_0$.
For more details, see pages 3-6 of ref.~\cite{mate-ell1}.

Table 1, Figures 1 and 2 indicate
that the Buda-Lund hydro model works well
at both RHIC energies:
it gives a good quality description of the
transverse mass dependence of the HBT radii as well as identified particle
spectra and elliptic flow,
see refs.~\cite{Csorgo:1995bi,Csorgo:2002ry,nr-inf,relsol-ell} for further details.

The temperature was estimated to be above $T_c$
in the most central 1/8th of the expanding ellipsoid~\cite{mate-ell1},
similarly to a fireball that is heated from inside.
We interpret this result as a
confirmation of the quark deconfinement and the cross-over like transition found in
Buda-Lund fits to central Au+Au collisions at
$\sqrt{s_{\NN}} = 130$ and $200$ GeV.

\begin{table}[ht]
\begin{center}
\begin{tabular}{|l|rl|rl|rl|rl|}
\hline
\hline
Buda-Lund parameter
		& \multicolumn{2}{c|}{Au+Au \@ 200 GeV}
		& \multicolumn{2}{c|}{Au+Au \@ 130 GeV}
\\
		\hline
		\hline
$T_0$ [MeV]           
		& \hspace{0.3cm} 196    &$\pm$ 13 
		& \hspace{0.3cm} 214    &$\pm$ 7
		\\
$T_{\mbox{\rm e}}$ [MeV]
		& 117    &$\pm$ 12
		& 102    &$\pm$ 11
		\\
$\mu_B$ [MeV]
		& 61	& $\pm$  52 
		& 77 	& $\pm$  38
		\\ \hline
$R_{G}$ [fm]          & 13.5   &$\pm$ 1.7
		& 28.0   &$\pm$ 5.5
		\\
$R_{s}$ [fm]          & 12.4    &$\pm$ 1.6
		& 8.6    &$\pm$ 0.4
		\\
$\langle u_t^\prime \rangle$
		& 1.6   &$\pm$  0.2
		& 1.0   &$\pm$  0.1
		\\ \hline
$\tau_0$ [fm/c]       & 5.8    &$\pm$ 0.3
		& 6.0    &$\pm$ 0.2
		\\
$\Delta\tau$ [fm/c]   & 0.9    &$\pm$ 1.2
		& 0.3    &$\pm$ 1.2
		\\
$\Delta\eta$          & 3.1    &$\pm$ 0.1
		& 2.4    &$\pm$ 0.1
		\\ \hline
$\chi^2/\mbox{\rm NDF}$
		& 114    &/ 208
		& 158.2  &/ 180
		\\ \hline \hline
\end{tabular}
\caption
{Buda-Lund hydro model v1.5 source parameters, corresponding to Figs. 1 and 2. 
The errors on the parameters are preliminary, as point-to-point and normalization errors are added in quadrature when evaluating $\chi^2$.
}
\label{tab:results}
\end{center}
\end{table}
%\section{Conclusions}

This work was supported by  following grants:
OTKA T034269, T038406, the OTKA-MTA-NSF grant INT0089462, the
NATO PST.CLG.980086 grant and the exchange program
of the Hungarian and Polish Academy of Sciences.


\begin{thebibliography}{99}
\vspace{3mm}
\bibitem{Csorgo:1995bi}
T.~Cs\"org\H{o} and B.~L\"orstad,
%``Bose-Einstein Correlations for Three-Dimensionally Expanding, Cylindrically Symmetric, Finite Systems,
Phys.\ Rev.\ C {\bf 54} (1996) 1390
%[arXiv:hep-ph/9509213].
%%CITATION = HEP-PH 9509213;%%

\bibitem{ster-ismd03}
M. Csan\'ad, T. Cs\"org\H{o}, B. L\"orstad, A. Ster,
%``An indication for deconfinement in Au + Au collisions at RHIC,''
Act. Phys. Pol. B{\bf 35}, 191 (2004),
nucl-th/0311102.
%%CITATION = NUCL-TH 0311102;%%

\bibitem{mate-warsaw03}
M.~Csan\'ad, T.~Cs\"org\H{o}, B.~L\"orstad and A.~Ster,
%``A hint at quark deconfinement in 200-GeV Au + Au data at RHIC,''
nucl-th/0402037.
%%CITATION = NUCL-TH 0402037;%%

\bibitem{Fodor:2001pe}
Z.~Fodor and S.~D.~Katz,
%``Lattice determination of the critical point of QCD at finite T and mu,''
JHEP {\bf 0203} (2002) 014
%[arXiv:hep-lat/0106002].
%%CITATION = HEP-LAT 0106002;%%


\bibitem{mate-ell1}
%\bibitem{Csanad:2003qa}
M.~Csan\'ad, T.~Cs\"org\H{o} and B.~L\"orstad,
%``Buda-Lund hydro model for ellipsoidally symmetric fireballs and the elliptic
%flow at RHIC,''
nucl-th/0310040.
%%CITATION = NUCL-TH 0310040;%%

\bibitem{ster-qm99}
A.~Ster, T.~Cs\"org\H{o} and B.~L\"orstad,
%``The reconstructed final state of Pb + Pb 158-GeV/A reactions from  spectra
%and correlation data of NA49, NA44 and WA98,''
Nucl.\ Phys.\ A {\bf 661}, 419 (1999)
hep-ph/9907338.
%%CITATION = HEP-PH 9907338;%%

\bibitem{cs-rev}
%\bibitem{Cs\"org\H{o}:1999sj}
T.~Cs\"org\H{o},
%``Particle interferometry from 40-MeV to 40-TeV,''
Heavy Ion Phys.\  {\bf 15}, 1 (2002)
%[arXiv:hep-ph/0001233].
%%CITATION = HEP-PH 0001233;%%

\bibitem{PHENIX-v2-id}
%\bibitem{Adler:2003kt}
S.~S.~Adler {\it et al.}  [PHENIX Collaboration],
%``Elliptic flow of identified hadrons in Au + Au collisions at  s(NN)**(1/2) = 200-GeV,
nucl-ex/0305013.
%[Phys.\ Rev.\ Lett.\  in press]
%%CITATION = NUCL-EX 0305013;%%

\bibitem{PHOBOS-v2} B.~B.~Back {\it et al.}  [PHOBOS Coll.],
%``Pseudorapidity and centrality dependence of the collective flow of  charged particles in Au + Au collisions at s(NN)**(1/2) = 130-GeV,
Phys.\ Rev.\ Lett.\  {\bf 89}, 222301 (2002)
%    [arXiv:nucl-ex/0205021].
%%CITATION = NUCL-EX 0205021;%%

\bibitem{PHOBOS-v2-qm02}
S.~Manly {\it et al.}  [PHOBOS Collaboration],
%``Flow and Bose-Einstein correlations in Au - Au collisions at RHIC,''
Nucl.\ Phys.\ A{\bf 715}, 611 (2003)
%[arXiv:nucl-ex/0210036].
%%CITATION = NUCL-EX 0210036;%%


\bibitem{Bearden:2001xw}
I.~G.~Bearden {\it et al.}  [BRAHMS Collaborations],
%``Charged particle densities from Au + Au collisions at s(NN)**(1/2) = 130-GeV,
Phys.\ Lett.\ B {\bf 523} (2001) 227
%[arXiv:nucl-ex/0108016].
%%CITATION = NUCL-EX 0108016;%%

\bibitem{Adcox:2001mf}
K.~Adcox {\it et al.}  [PHENIX Collaboration],
%``Centrality dependence of pi+-, K+-, p and anti-p production from s(NN)**(1/2) = 130-GeV Au + Au collisions at RHIC,
Phys.\ Rev.\ Lett.\  {\bf 88} (2002) 242301
%[arXiv:nucl-ex/0112006].
%%CITATION = NUCL-EX 0112006;%%

\bibitem{Adcox:2002uc}
K.~Adcox {\it et al.}  [PHENIX Collaboration],
%``Transverse mass dependence of two-pion correlations in Au + Au  collisions at s(NN)**(1/2) = 130-GeV,
Phys.\ Rev.\ Lett.\  {\bf 88} (2002) 192302
%[arXiv:nucl-ex/0201008].
%%CITATION = NUCL-EX 0201008;%%

\bibitem{Back:2001bq}
B.~B.~Back {\it et al.}  [PHOBOS Collaboration],
%``Charged-particle pseudorapidity density distributions from Au + Au  collisions at s(NN)**(1/2) = 130-GeV,
Phys.\ Rev.\ Lett.\  {\bf 87} (2001) 102303
%[arXiv:nucl-ex/0106006].
%%CITATION = NUCL-EX 0106006;%%


\bibitem{Adler:2001zd}
C.~Adler {\it et al.}  [STAR Collaboration],
%``Pion interferometry of s(NN)**(1/2) = 130-GeV Au + Au collisions at  RHIC,
Phys.\ Rev.\ Lett.\  {\bf 87} (2001) 082301
%[arXiv:nucl-ex/0107008].
%%CITATION = NUCL-EX 0107008;%%

\bibitem{Bearden:2001qq}
I.~G.~Bearden {\it et al.}  [BRAHMS Collaboration],
%``Pseudorapidity distributions of charged particles from Au + Au  collisions at the maximum RHIC energy,
Phys.\ Rev.\ Lett.\  {\bf 88} (2002) 202301
%[arXiv:nucl-ex/0112001].
%%CITATION = NUCL-EX 0112001;%%

\bibitem{Adler:2003cb}
S.~S.~Adler {\it et al.}  [PHENIX Collaboration],
%``Identified charged particle spectra and yields in Au + Au collisions at
%s(NN)**(1/2) = 200-GeV,''
nucl-ex/0307022.
%%CITATION = NUCL-EX 0307022;%%

\bibitem{Adler:2004rq}
S.~S.~Adler {\it et al.} [PHENIX Collaboration],
%``Bose-Einstein correlations of charged pion pairs in Au + Au collisions at s(NN)**(1/2) = 200-GeV,''
nucl-ex/0401003.
%%CITATION = NUCL-EX 0401003;%%

\bibitem{Csorgo-blhome}
{http://www.kfki.hu/$\tilde{\,\,\,}$csorgo/budalund/}

\bibitem{blastwave}
%\cite{Retiere:2003kf}
%\bibitem{Retiere:2003kf}
F.~Retiere and M.~A.~Lisa,
 %``Observable implications of geometrical and dynamical aspects of freeze-out
%in heavy ion collisions,''
arXiv:nucl-th/0312024.
%%CITATION = NUCL-TH 0312024;%%



\bibitem{cracow-sp}
%\cite{Broniowski:2001we}
%\bibitem{Broniowski:2001we}
W.~Broniowski and W.~Florkowski,
%``Explanation of the RHIC p(T)-spectra in a thermal model with expansion,''
Phys.\ Rev.\ Lett.\  {\bf 87} (2001) 272302
[arXiv:nucl-th/0106050].
%%CITATION = NUCL-TH 0106050;%%

\bibitem{cracow-hbt}
%\cite{Broniowski:2002wp}
%\bibitem{Broniowski:2002wp}
W.~Broniowski, A.~Baran and W.~Florkowski,
%``Thermal model at RHIC. II: Elliptic flow and HBT radii,''
AIP Conf.\ Proc.\  {\bf 660} (2003) 185
%[arXiv:nucl-th/0212053].
%%CITATION = NUCL-TH 0212053;%%




\bibitem{Csorgo:2002ry}
T.~Cs\"org\H o and A.~Ster,
%``The reconstructed final state of Au + Au collisions from PHENIX and  STAR data at s**(1/2) = 130-A-GeV: Indication for quark deconfinement at RHIC,
Heavy  Ion Phys.\  {\bf 17} (2003) 295
%[arXiv:nucl-th/0207016].
%%CITATION = NUCL-TH 0207016;%%



\bibitem{nr-inf}
%\bibitem{Cs\"org\H{o}:2002kt}
T.~Cs\"org\H{o} and J.~Zim\'anyi,
%``Inflation of fireballs, the gluon wind and the homogeneity of the HBT  radii at RHIC,
Heavy Ion Phys.\  {\bf 17}, 281 (2003)
%[arXiv:nucl-th/0206051].
%%CITATION = NUCL-TH 0206051;%%


\bibitem{relsol-ell}
%\bibitem{Cs\"org\H{o}:2003ry}
T.~Cs\"org\H{o}, L.~Csernai, Y.~Hama, T.~Kodama,
%``Simple solutions of relativistic hydrodynamics for systems with  ellipsoidal symmetry,
nucl-th/0306004.
%%CITATION = NUCL-TH 0306004;%%

\end{thebibliography}
\end{document}